
%
%
%
%
%
%

\input phyzzx
\nonstopmode
\twelvepoint
\nopubblock
\overfullrule=0pt
\tolerance=5000

\line{\hfill }
\line{\hfill PUPT-1389, IASSNS-HEP-93/16}
\line{\hfill March 1993}

\titlepage
\title{Emergence of Coherent Long Wavelength Oscillations After
a Quench: Application to QCD}

\author{Krishna Rajagopal\foot{Research supported in part by a Natural
Sciences and Engineering Research Council of Canada 1967 Fellowship
and in part by a Charlotte Elizabeth Procter Fellowship.~~~
RAJAGOPAL@PUPGG.PRINCETON.EDU}}
\vskip .2cm
\centerline{{\it Department of Physics }}
\centerline{{\it Joseph Henry Laboratories }}
\centerline{{\it Princeton University }}
\centerline{{\it Princeton, N.J. 08544 }}

\author{Frank Wilczek\foot{Research supported in part by DOE grant
DE-FG02-90ER40542.~~~WILCZEK@IASSNS.BITNET}}
\vskip.2cm
\centerline{{\it School of Natural Sciences}}
\centerline{{\it Institute for Advanced Study}}
\centerline{{\it Olden Lane}}
\centerline{{\it Princeton, N.J. 08540}}
\endpage

\abstract{To model the dynamics of
the chiral order parameter in a far from equilibrium
phase transition, we consider quenching in the O(4) linear
sigma model.   We argue, and present
numerical evidence,
that in the period immediately following the quench
long wavelength modes of the pion field are amplified.
This results in large regions of coherent pion oscillations, and
could lead to
dramatic phenomenological
consequences in heavy ion collisions.}

\endpage

\REF\anselm{A. Anselm and M. Ryskin, {\it Phys. Letters} {\bf B226},
482 (1991).}

\REF\blaizot{J.-P. Blaizot and A. Krzywicki, {\it Phys. Rev.} {\bf D46}, 246
(1992).}

\REF\recentbjorken{J. D Bjorken, {\it Int. J. Mod. Phys.} {\bf A7}, 4189
(1992);
J. D. Bjorken, {\it Acta Physica Polonica\/} {\bf B23}, 561 (1992).}

\REF\kowalski{K. L. Kowalski and C. C. Taylor, {\it Disoriented Chiral
Condensate:  A White Paper for the Full Acceptance Detector}, CWRUTH-92-6,
hep-ph/9211282, 1992.}

\REF\us{K. Rajagopal and F. Wilczek, {\it
Static and Dynamic Critical Phenomena at a Second
Order QCD Phase Transition\/ } PUPT-1347, IASSNS-HEP-92/60,
hep-ph/9210253, to appear in {\it Nucl. Phys.}
{\bf B}.}

\REF\centauro{C. M. G. Lattes, Y. Fujimoto, and S. Hasegawa, {\it Phys. Rept.}
{\bf 65}, 151 (1980).}

\REF\jacee{J. J. Lord and J. Iwai, Univ. of Washington preprint
(paper 515 submitted to the International Conference on High Energy
Physics, Dallas, August 1992); J. Iwai (JACEE collaboration), UWSEA
92-06.}

\REF\wilczek{F. Wilczek, {\it Int. J. Mod. Phys.} {\bf A7}, 3911 (1992).
This elaborates earlier work of R. Pisarski and F. Wilczek, {\it Phys. Rev.}
{\bf D29}, 338 (1984).}


\REF\gm{M. Gell-Mann and M. Levy, {\it Nuovo Cimento} {\bf 16}, 705 (1960).}

\REF\bray{A. J. Bray, {\it Phys. Rev.} {\bf B41}, 6724 (1990);  T. J. Newman,
A. J. Bray, and M. A. Moore, {\it Phys. Rev.} {\bf B42}, 4514 (1990); and
references therein.}

\REF\bjorken{J. D. Bjorken, {\it Phys. Rev.} {\bf D27}, 140 (1983).}

\REF\turgel{N. Turok and D. N. Spergel, {\it Phys. Rev. Lett.} {\bf 66}, 3093
(1991).}

\REF\press{D. N. Spergel, N. Turok, W. H. Press, and B. S. Ryden,
{\it Phys. Rev.} {\bf D43}, 1038 (1991).}

\REF\ryden{W. H. Press, B. Ryden, and D. N. Spergel, {\it Astrophys. J.}
{\bf 347}, 590 (1989).}

\REF\doljac{L. Dolan and R. Jackiw, {\it Phys. Rev.} {\bf D9}, 3320 (1974).}

\REF\loewe{C. Contreras and M. Loewe, {\it Int. J. Mod. Phys.} {\bf A5},
2297 (1990).}

\REF\bernard{C. Bernard {\it et al.}, {\it Phys. Rev.} {\bf D45},
3854 (1992).}

Among the most interesting speculations regarding
ultra-high energy hadronic or heavy nucleus collisions is the idea
that regions of misaligned vacuum might occur
[\anselm ,\blaizot ,\recentbjorken ,\kowalski ,\us ].
In such misaligned
regions, which are analogous to misaligned domains in a ferromagnet,
the chiral condensate points in a different direction from that
favored in the ground state.  If we parametrize the condensate
using the usual variables of the sigma model, misaligned vacuum regions
are places where the four-component
$(\sigma, \vec \pi) $ field, that in the ground state takes
the value $(v,0)$, is partially aligned in the
$\pi$ directions.  If they were produced,
misaligned vacuum regions plausibly would
behave as ``pion lasers'', relaxing to the ground state by coherent
pion emission.  They would produce clusters of pions bunched in rapidity
with highly non-Gaussian charge distributions.  Thus a misaligned
vacuum region starting with the field in the $\pi_1 - \pi_2$ plane would
emit only charged pions (equally positive and negative, since the
fields are real), while a misaligned vacuum region starting with the
field pointing
in the $\pi_3$ direction would emit only neutral pions.
More generally if we define
$$
R \equiv { n_{\pi^0} \over n_{\pi^0} + n_{\pi^+ \pi^-} }
\eqn\jc
$$
then
assuming that all initial values on the 3-sphere are
equally probable,
the probability distribution ${\cal P}(R)$ is given by
$$
{\cal P}(R) = {1 \over 2} R^{-1/2} ~.
\eqn\je
$$

As an example of \je , we note that the probability that the
neutral pion fraction $R$ is less than .01 is 0.1!  This is a
graphic illustration of how different \je\ is from what one would
expect if individual pions were independently emitted with no
isospin correlations.  According to
\je\ there is a substantial probability that,
say, a cluster of 70 pions would all be charged -- something that
could essentially never occur for incoherent emission.  There may
be some hint of such behavior in the Centauro
(overwhelmingly charged)
and anti-Centauro
(overwhelmingly neutral) events reported in the
cosmic ray literature.[\centauro ,\jacee ]~
Here
we propose a concrete mechanism by which such phenomena
might arise in heavy ion
collisions, for which the plasma
is far from thermal equilibrium.

In previous work [\us ,\wilczek ]
we considered the equilibrium phase structure of
QCD.  We argued that QCD with two massless quark flavors probably undergoes
a second-order chiral phase transition.  For many purposes it is a
good approximation to treat the u and d quarks as approximately massless,
and we expect that real QCD has a smooth but perhaps rapid transition as
a function of decreasing
temperature from small intrinsic to large spontaneous chiral
symmetry breaking.  At first sight it might appear that a second-order
phase transition is especially favorable for the development of large
regions of misaligned vacuum.  Indeed the long-lived, long-wavelength
critical fluctuations which provide the classic signature of a second-order
transition {\it are\/} such regions.   Unfortunately the effect of the
light quark masses, even though they are formally much smaller than
intrinsic QCD scales, spoil this possibility [\us ].  The pion masses,
or more precisely the inverse correlation length in the pion channel, are
almost certainly {\it not\/} small compared to the transition temperature.
As a result the misaligned regions are modest affairs at best
even near the critical temperature.  They
extend at most over a few thermal
wavelengths, and almost certainly do not contain sufficient energy to
radiate
large numbers of correlated pions.

Here we will consider an idealization that is in
some ways opposite to that of thermal equilibrium, that is the
occurence of a sudden
quench from high to low temperatures, in which the
$(\sigma ,\vec \pi)$ fields are suddenly removed from contact with a
high temperature heat bath and subsequently evolve mechanically.
We shall show that long wavelength fluctuations of the
light fields (the pions in QCD, which would be massless if not
for the quark masses) can develop following a quench from
some temperature $T~ >~ T_c$ to $T~=~0$.
The long wavelength modes of the pion fields are unstable and
grow relative to the short wavelength modes.

Before we enter into the details of our model and simulations it
seems appropriate to discuss the qualitative reason for the result
just mentioned, which we suspect may be of wider interest.
Whenever one has spontaneous breaking of a
continuous global symmetry, massless
Nambu-Goldstone bosons occur.  The masslessness of these modes occurs
through a cancellation following the schema
$$
m^2~=~ -\mu^2 ~+~ \lambda \phi^2~,
\eqn\cancel
$$
where the second term arises from interaction with a condensate whose
expectation value $\langle \phi \rangle^2$
satisfies $\langle \phi \rangle^2~=~\mu^2/\lambda~\equiv~v^2$
in the ground
state.  However following
a quench the condensate starts with its average
at $0$ rather than $v$,
and generally oscillates before settling to its
final value.  Whenever $~\langle \phi \rangle ^2 < v^2$, $m^2$ will be
negative, and {\it sufficiently long wavelengths fluctuations in the
Nambu-Goldstone
boson field will grow exponentially}.
The same mechanism will work, though
less efficiently, if one has only an approximate symmetry and approximate
Nambu-Goldstone modes, or for that matter
other modes whose effective $m^2$ is
`accidentally' pumped negative by their interactions with the condensate.

Returning to QCD, we shall use the classic
Gell-Mann--Levy Lagrangian [\gm ] to describe the
low energy interactions of pions:
$$
{\cal L}  = \int d^4 x \Bigl\lbrace {1\over 2}~\partial^i \phi^\alpha
     \partial_i \phi_\alpha ~-~{\lambda \over 4}
     \bigl( \phi^\alpha \phi_\alpha ~-~ v^2 \bigr) ^2 ~+~
      H \sigma ~\Bigr\rbrace ~,
\eqn\bb
$$
where $\phi$ is a four-component vector in internal space
with entries $(\sigma , \pi )$.
Here, $\lambda$, $v$, and $H\propto m_q$ are to be thought of as
parameters in the low energy effective theory obtained after integrating
out heavy degrees of freedom.
We shall treat \bb\ as it stands as a classical field theory, since
the phenomenon we are attempting to address is basically classical and
because as a practical matter it would be prohibitively difficult to
do better.  We shall be dealing with temperatures of order 200 MeV or
less, so that neglect of heavier fields seems reasonable.

To model a quench, we begin
at a temperature well above $T_c$.  The typical
configurations have short correlation lengths
and $\langle \phi \rangle \sim 0$.  ($\langle \phi \rangle
\neq 0$ because $H \neq 0$.)  One then takes the temperature
instantaneously to zero.  The equilibrium configuration
is an ordered state with the $\phi$ field aligned in the $\sigma$
direction throughout space, but this is not the configuration
in which the system finds itself.  The actual, disordered
configuration then
evolves
according to the zero temperature equations of motion obtained
by varying \bb .  Quenching in magnet models has been much studied
in condensed matter physics [\bray ].  However in that context it
is usually
appropriate to use diffusive equations of motion, because the magnet is
always in significant contact with other light modes (\eg\ phonons).
For this reason the condensed matter literature we are aware of
does not directly apply
to our problem.

Let us now consider how the physics of quenching may be applied in
relativistic heavy ion collisions.  In Bjorken's [\bjorken ] picture
of such a collision, the incident nuclei as seen in the center of
mass frame are both Lorentz contracted into pancake shapes.  They
pass through each other, and leave behind a region of hot vacuum.
The baryon number of the incident nuclei ends up in that part of the
plasma heading approximately down the beam pipe, and the central
rapidity region which we consider in this paper consists of plasma
with approximately zero baryon number which expands and cools
through $T=T_c$ and eventually hadronizes into the detected pions.
If the collision is energetic enough
to create a region of plasma well above $T=T_c$, the $\phi$
field will indeed be fluctuating among an ensemble of disordered
configurations.
The plasma cools rapidly as it expands, and cannot
be exactly in thermal equilibrium.  If it cools fast enough, the
configuration  of the $\phi$ field will ``lag,'' and as in a quench
the system will find itself more disordered than
the equilibrium configuration appropriate to the current temperature.

We have done  numerical simulations of quenching to zero temperature
in the linear sigma model with an explicit symmetry breaking term $H \sigma$
which makes the pions massive.
Turok and Spergel [\turgel ] have considered
this scenario with no explicit symmetry breaking term as a cosmological
model for large scale structure formation in the early universe.
They find a scaling solution in which the size of correlated
domains grows at the speed of light as larger and larger regions
come into causal contact and align.  When the
$O(N)$ symmetry
is explicitly broken, however, we do not expect a scaling solution.
The $H \sigma$ term tilts the potential,
the vacuum manifold is not
degenerate,
and in a time of order $m_\pi^{-1}$ the scalar field
in all regions (whether in causal contact or not) will be oscillating
about the sigma direction.

In numerically simulating a quench, we choose $\phi$ and $\dot \phi$
randomly independently on each site of a cubic lattice.  This means
that the lattice spacing $a$ represents the correlation length
in the initial conditions.  If the initial conditions are chosen
to model a thermal ensemble at some temperature $T>T_c$, then
the $O(N)$ symmetry is not spontaneously broken, and the lattice
spacing $a$ represents
the $\pi$ and
$\sigma$ correlation lengths which are approximately
degenerate.
Because the initial conditions must not distinguish
between the $\sigma$ and $\pi$ directions in internal
space, we must use the linear sigma model \bb\ instead of
integrating out the $\sigma$ which is heavy at zero
temperature to obtain the nonlinear sigma model.
In the simulation whose results are shown
in Figure~1, we chose $\phi$ and $\dot \phi$ randomly from gaussian
distributions centered around $\phi = \dot \phi =0$ and with
root mean square variance $v/2$ and $v$ respectively.
The three parameters $v$, $H$, and $\lambda$ in \bb\
determine $m_\pi$, $m_\sigma$, and $f_\pi=\langle 0|\sigma |0\rangle$
according to
$$
\lambda \langle 0 | \sigma |0 \rangle \Bigr(
\langle 0 | \sigma |0 \rangle^2 ~-~v^2 \Bigr) ~-~ H ~=~0~~,
\eqn\bc
$$
$$
m_\pi^2={H \over  \langle 0 | \sigma |0 \rangle }~~,~~~~{\rm and}~~~~
m_\sigma^2=3\lambda  \langle 0 | \sigma |0 \rangle^2 - \lambda v^2~~.
\eqn\bd
$$
Note that $f_\pi = \langle 0| \sigma |0 \rangle > v$ for $H\neq 0$.
In interpreting our results we must remember that
while in the code we are free to choose the energy scale
by setting the lattice spacing $a=1$,
$a$ actually represents the initial correlation length.
In choosing
the parameters for Figure~1 we
assumed that $a=(200~{\rm MeV})^{-1}$ and then chose
$v=87.4 ~{\rm MeV} = 0.4372~a^{-1}$,
$H=(119 ~{\rm MeV})^3 = 0.2107~a^{-3}$,
and $\lambda = 20.0$ so that $f_\pi=92.5~ {\rm MeV}$,
$m_\pi=135~ {\rm MeV}$,
and $m_\sigma = 600 ~{\rm MeV}$.

With parameters and initial conditions chosen, we evolve
the initial configuration according to the equations of motion
using a standard finite difference,
staggered leapfrog scheme.  (For details of the numerical
method see [\press ,\ryden ] and in particular equations 32-35 of
[\ryden ].)  We used a time step $dt = {\rm min}( {a \over 10},
{2 \pi \over 10 m_\sigma })$.  We verified that our results do not change
if the time step is reduced.
After each two time steps,
we computed the spatial fourier transform of the configuration,
and from that obtained the angular averaged power spectrum.
In Figure~1,
we plot the power in modes of the pion and sigma fields
with spatial wave vectors of
several different magnitudes $k$ as a function of time.
All the curves start at approximately the same
value
at $t=0$ because the initial power spectrum is white
since we chose $\phi$ independently at each lattice site.
The behaviour of the low momentum pion modes is striking.
While the initial power spectrum is white and, as
ergodicity arguments would predict, the system at late times
is approaching an equilibrium configuration in which the
equipartition theorem holds, at intermediate times of order
several times $m_\pi^{-1}$ the low momentum pion modes are oscillating
coherently in phase with large amplitudes.

The simulation was done in a $64^3$ box.  We verified that
finite size effects are not important even for the longest
wavelength ($ka=0.20$) mode shown in Figure~1 by checking that the
behaviour of the $ka=0.31$ mode is the same in a $32^3$ and
a $64^3$ box.  The behaviour of only one component of the
pion field is shown in Figure~1a.  The other two look qualitatively
the same.  It should be noted that the exact height of the peaks
in the curves depend on the specific initial conditions.
If the simulation is run with the same initial distribution for
$\phi$ and $\dot \phi$ but with a different seed for the random
number generator, the heights of the peaks
change, and the relative sizes of the peaks in the three different
pion directions change.  The qualitative features of Figure~1 ---
the growth of long wavelength modes of the pion field --- do not depend
on the specific realization of the initial conditions.
We discuss below how
the results change as a function of parameters in the potential and
the initial distributions.

Let us compare the growth of long wavelength modes found numerically
with expectations from the mechanism previously discussed.
Suppose the potential $V$ in \bb\ were simply $V(\phi) = (m^2/2)\phi^\alpha
\phi_\alpha$.  Then, the equations of motion would be linear, and
modes with different spatial wave vector $\vec k$ would be uncoupled.
Each curve in Figure~1 would be a sinusoid with period $\pi / \sqrt{m^2 +
\vec k ^2}$ and constant amplitude.  (The power spectrum, being quadratic
in the fields, oscillates with one half the period of the fields.)
The period of the oscillations in Figure~1a is indeed given by
$\pi / \sqrt{m_\pi^2 + \vec k ^2}$, but the amplitudes are far from
constant.  This behaviour can be qualitatively understood by
approximating $\phi^2$ in the nonlinear term in the equation of motion
by its spatial average:
$$
\phi^\alpha \phi_\alpha (\vec x, t) \sim \langle \phi_\alpha \phi^\alpha
\rangle (t) ~~.
\eqn\bg
$$
This approximation is exact in the large $N$ limit [\turgel ].
In our problem,
$N=4$.
Using \bg\ and doing the spatial fourier transform, the equation
of motion for the pion field becomes
$$
{d^2 \over dt^2} \vec \pi (\vec k,t) =
- m_{eff}^2 (k,t) \vec \pi (\vec k,t)
\eqn\bgg
$$
with the time dependent ``mass'' given by
$$
m_{eff}^2 (k,t) \equiv - \lambda v^2 + k^2
+ \lambda \langle \phi^2 \rangle (t)
\eqn\bh
$$
where $k=|\vec k|$.

Figure~2 shows the time evolution of $\langle \phi^2 \rangle$ in the
same simulation whose results are shown in Figure~1.  In the
initial conditions, $\phi$ is gaussian distributed with
$\langle \phi^2 \rangle < v^2$.
Therefore, for a range of wavevectors with $k$ less than
some critical value, $m_{eff}^2 < 0$ and the long wavelength
modes of the pion field start growing exponentially.
$\langle \phi^2
\rangle$ grows and then executes damped oscillations
about its ground state value
$\langle 0|\sigma |0 \rangle ^2$.
A wave vector $k$ mode of the pion field is
unstable and grows exponentially whenever $\langle \phi^2 \rangle <
v^2 - k^2/\lambda$.  Modes with $k^2 > \lambda v^2$ can never be unstable.
The $k=0$ mode is unstable during
the periods of time when the $\langle \phi^2
\rangle$ curve in Figure~2 is below $v^2$.
Since $\langle \phi^2 \rangle$ is
oscillating about
$\langle 0|\sigma |0 \rangle^2 > v^2$, after some time the
oscillations have damped enough that $\langle \phi^2 \rangle$ never drops
below $v^2$, and from that time on all modes
are always stable and oscillatory.
In general, longer wavelength modes are unstable
for more and for longer
intervals of time than shorter wavelength modes
as $\langle \phi^2 \rangle$ oscillates.
Also, $m_{eff}^2$ is more negative and more growth occurs for
modes with smaller $k$.
Making the approximation \bg\ and thus using \bgg\
cannot be expected to completely reproduce the effects
of the nonlinear term in \bb\ which is local in position space.
Nevertheless, it predicts that the long wavelength modes of
the pion field go through alternating periods of
oscillatory behaviour and exponential growth spurts and therefore
gives us a good understanding of the behaviour of these modes
in Figure~1a.
Because
the timing of the growth spurts for different (low) $k$ modes are all
determined by $\langle \phi^2 \rangle$, different modes have their
growth spurts at the same times and all the long wavelength modes in
Figure~1a oscillate in phase.
The longer wavelength modes have more, stronger, and longer
growth spurts,
and are therefore amplified more as in Figure~1a.
In an equilibrium phase transition, explicit symmetry breaking
keeps the correlation length at $T_c$ finite and,
in the case of QCD, too short to be of interest.[\us ]
After a quench, on the other hand,
arbitrarily long
wavelength modes of the pion field are amplified even though
the pion mass is non-zero.

What about the $\sigma$ modes?  For $\sigma^2 < v^2 /3$, the effective
${\rm mass}^2$ for the $k=0$ mode is negative.  This condition
is far less likely to be satisfied than $\langle \phi^2 \rangle < v^2$.
Hence, while some growth of the
low momentum $\sigma$ modes is
possible, as Figure~1 confirms, they do not grow as much as the
low momentum pion modes do.  The behaviour of the sigma modes is also
complicated by the fact that oscillations with two periods ---
$2 \pi / m_\pi$ and $2 \pi / m_\sigma$ --- contribute. Roughly, this occurs
because oscillations
ostensibly along the pion directions are oscillations in a
curved valley and are therefore seen in sigma also.

At late times $\langle \phi^2 \rangle$
oscillates with small enough amplitude that it never
goes below $v^2$ and consequently no modes are ever unstable.
Hence, if we make the approximation \bg\ we would expect that at
late times each mode in Figure~1a would continue to oscillate with
approximately constant amplitude, with the longer wavelength modes
maintaining the large amplitudes acquired during their exponential
growth spurts.
That this is not what is seen in Figure~1a
reflects the effects we neglected in making the approximation \bg .
Because
the modes are in fact coupled and the equations of motion are
actually nonlinear, ergodicity arguments suggest that eventually
equipartition should apply.  If the energy is equally divided among
modes, then the $({\rm amplitude})^2$ in a mode should be inversely
proportional to $(m_\pi^2 + k^2)$.  Also, the amplitude of the
sigma modes at late times should be less than that of pion modes
of the same $k$ because $m_\sigma > m_\pi$.
The results shown in Figure~1
are consistent with the assumption that at late times equipartition
applies, although at $t=80a$ the longest wavelength pion modes are
still decreasing in amplitude.  (We have verified that at even
later times they do reach equipartition.)
It is reasonable that longer wavelength modes take
longer to decrease in amplitude than shorter wavelength modes.
As the system heads toward equipartition, power has to flow from
low momentum modes to higher momentum modes.
This plausibly
happens more quickly at higher momentum because higher momentum
modes are coupled to a larger number of modes, or, in the continuum,
a larger volume of momentum space.
Hence, we end up with the striking phenomena of Figure~1a:
while the system begins and ends  with
small $\vec \pi (\vec k,t)$, in between the long wavelength pion modes
are dramatically amplified.

In modelling a heavy ion collision, we should include the effects
of the expansion of the plasma after a quench.  In an equilibrium
phase transition, the expansion causes a decrease in the temperature.
After a quench, however, $T=0$ and the expansion has the effect of
reducing the energy per unit volume in the system.
The simplest way to model this is to introduce a time dependent
lattice spacing $a(t)$.  This introduces a $\dot \phi \dot a /a$ term
in the equations of motion which, for $a(t)$ increasing with time,
leads to damping.  Doing this, we found that each of the
modes (now defined by spatially fourier transforming from
comoving $\vec x$ to comoving $\vec k$) red shifts with time.
However, as long as the $\dot \phi$ term does not dominate
over the $\ddot  \phi$ term low momentum modes still become
unstable and grow with respect to high momentum modes.
In a real heavy ion collision, the expansion is anisotropic.
The plasma expands much more rapidly along the beam direction
than in the transverse directions [\bjorken ].
While we have
not attempted to model this explicitly, it probably
means that correlated volumes do not grow large in the
longitudinal direction.  The large expansion rate will
rapidly damp modes of the pion field with wave vectors in the
longitudinal direction.
However, it seems possible that
for a particular longitudinal position (or, equivalently [\bjorken ],
a particular rapidity) coherent fluctuations on length scales
all the way up to the transverse extent of the plasma could develop.
Another effect of the expansion is that after some time
(perhaps of order 10 fm$^{-1} \sim 7 m_\pi^{-1}$ [\bjorken ])
the energy density
is low enough that the description in terms of classical fields
no longer makes sense.  After this time, one has individual
pions flying off towards the detector.  From Figure~1a, it is plausible
that at this time there are large long wavelength oscillations of
the pion field.

If the initial conditions
or parameters are such that the initial value of the mean
energy per unit volume $E$ is much more than ${\lambda \over 4} v^4$
no striking growth of low momentum modes occurs.
Indeed when the total energy is large,
the field will be able to climb far up the potential to $|\phi |\gg v$,
and therefore $\langle \phi^2 \rangle (t)$ will always be greater than
$v^2$, and no modes will become unstable.
Requiring the energy not to be too large compared to the height of the
potential at $\phi = 0$
can be seen either as a constraint on the parameters of the
potential ($V(\phi=0)\sim {\lambda \over 4} v^4$)
for given initial conditions or on
the initial conditions for a given potential.
In the simulation of Figure~1, the total energy
$E \sim 2 {\lambda \over 4} v^4$.
As parameters (or initial conditions) are changed so that
the total energy is increased, the growth of low momentum modes
becomes less and less prominent, and when
$E \gsim 10 {\lambda \over 4} v^4$,
dramatic growth does
not occur.

As long as $\phi$ is centered around
$\phi=0$ initially, and as long as $E$ is not too large,
the choice of initial distributions has little qualitative
effect.
We have tried distributions ranging from $\phi \equiv 0$
with $\dot \phi$ chosen from a gaussian distribution to
$\dot \phi \equiv 0$ with $\phi$ chosen from a uniform distribution
with $\phi^2 \leq v^2$, and in all cases obtained results similar
to those of Figure~1.

Let us attempt to estimate the conditions that might govern
a realistic quench.
We will make a crude estimate of the lattice spacing $a$
and initial distributions for $\phi$ and $\dot \phi$ using
the one loop temperature dependent effective potential with
$H=0$.  At large temperatures, this is given by [\doljac ,\loewe ]
$$
V(\phi ,T) = m^2(T) \phi^2 + {\lambda \over 4} \phi^4
\eqn\bi$$
where the temperature dependent mass is given by
$$
m^2(T) = \lambda ( -v^2 + T^2 /4 ) ~~.
\eqn\bj
$$
While this is a crude approximation, it gives
$T_c = 2v = 175 ~{\rm MeV}$ which is coincidentally quite close to the
lattice gauge theory value [\bernard ].
If we quench from an initial temperature $T_i$,
the lattice spacing will be given by the correlation
length at $T_i$
$$
a = m^{-1}(T_i)
\eqn\bk
$$
For $T_i = 1.2~T_c = 2.4~v$, this gives $a^{-1}=2.97~v = 260~{\rm MeV}$.

We will model initial conditions with correlation length $a$ by choosing
$\phi$ (and $\dot \phi$) independently at each lattice site.
$\phi$ should therefore be distributed according to the probability
distribution
$$
\exp - {a^3 \over T_i}
\Bigl( {1 \over 2} m^2(T_i) \phi^2 + {\lambda \over 4} \phi^4 \Bigr)~~.
\eqn\bl
$$
Because it is cut off by the quartic term, this distribution has much
less weight at large $|\phi|$ than a gaussian.
We approximate it by the truncated gaussian distribution
$$
\exp - \Bigl( {1  \over 2} m^2(T_i) \phi^2 \Bigr)~
{\rm for}~ |\phi|<\phi_{max}
{\rm ~;~~~} 0 ~{\rm for}~ |\phi|>\phi_{max}
\eqn\bm
$$
where $\phi_{max}$ is chosen so that $\phi^2$ has the same expectation
value in the distributions \bl\ and \bm .  To choose initial conditions
for $\dot \phi$, we note that $\langle \dot \phi ^2 \rangle =
\langle \phi^2 \rangle$ in lattice units.  (This is true for a quadratic
potential, and is also true in general if equipartition applies.)
Hence, we
use the same distribution for $\dot \phi$ as for $\phi$.
We now have a crude recipe for choosing initial conditions appropriate
to model a quench from an initial temperature $T_i$ to $T=0$.

What does our crude recipe predict? From \bl , we expect that
in lattice units $\phi_{max} \sim (\lambda /2)^{1/4} = 0.56$.
In fact, for $T_i=1.2~T_c$, we obtain $\phi_{max}=0.63~a^{-1}=1.87~v$.
The value of $\phi_{max}$ is crucial, because
it determines over how much of the potential $\phi$ is distributed.
We see that choosing $\phi$ gaussian distributed with variance
$v/2$ as we did in Figure~1
was overly optimistic.
For $T_i=1.2~T_c$, it is more reasonable to assume that
$\phi$ is distributed according to \bm\ with $\phi_{max}=1.87~v$.
With these initial conditions, the energy per unit volume turns
out to be $E\sim 8 {\lambda \over 4} v^2$
and the pion field behaves as
shown in Figure~3.  Low momentum modes do indeed grow, but
less dramatically than in Figure~1.  Also, the system finds its
way to a final state described by equipartition more quickly
than in Figure~1.  Nevertheless, there is a period of time after the
quench when low momentum modes have much more power than
higher momentum modes.

If $T_i$ is taken to be higher, $\phi_{\max} /v$ and $1/av$ increase,
$E/({\lambda \over 4} v^4)$ increases,
and the phenomenon of interest is washed out.
It is not unreasonable, however, to consider quenching from temperatures
larger than but comparable to $T_c$.  If the initial temperature
were much higher, the system could plausibly stay in thermal equilibrium
until close to $T_c$.  Hence, choosing $T_i = 1.2~T_c$ and thus obtaining
parameters and initial distributions like those of Figure~3 may be a
reasonable if crude approximation to a real heavy ion collision.

We have made many idealizations
and approximations.  Nevertheless, it seems possible to us
that the essential qualitative feature of the phenomenon we
have elucidated --- long wavelength pion modes experiencing periods of
negative mass$^2$ and consequent growth following a quench --- could
occur in real heavy ion collisions.
Given the explicit symmetry breaking, one might have expected
the dynamics following a quench from
a ``generic'' initial state to be featureless.
The mechanism here
discussed
provides a robust counterexample that should  be applicable
in other contexts besides the QCD phase transition.  Examples could
include
the reheating of some inflationary universe
models and the quenching of spin systems which order at low temperatures,
for which the interactions of the spins with phonons are less
important than their interactions among themselves.
If a heavy ion collision is
energetic enough that there is a central rapidity region
of high energy density and low baryon number, and if such a
region cools rapidly enough that it can be modelled as a quench,
this will be detected by observing clusters of pions
of similar rapidity in which the the fraction of neutral pions is fixed.
This ratio will be different in different clusters and will
follow a distribution like \je .
We can think of no process besides a QCD phase transition
with the chiral order parameter far out of equilibrium
that could produce such a signature.

\ack{We are grateful to David Spergel and Neil Turok for giving
us the code created by the authors of [\turgel , \press , and \ryden ],
and to both of them and Ue-Li Pen for many very
helpful conversations.}

\vfill
\eject

\FIG\zza{Time evolution of the power spectrum of the
pion field and the sigma field
at several spatial
wavelengths after a quench at $t=0$.
$\phi$ and $\dot \phi$ were chosen at time $t=0$
independently
at each lattice site from gaussian distributions centered at
the origin as described in the text.
The parameters
of the potential were chosen so that for
a lattice spacing $a=(200{\rm ~MeV})^{-1}$, the physical values
$m_\pi =135~{\rm MeV}$, $m_\sigma =600~{\rm MeV}$, and
$f_\pi =92.5~{\rm MeV}$ are obtained.
The simulation
was performed on a $64^3$ lattice
and the time step
was $a/10$.  After every two time steps, the spatial
fourier transform of $\phi$ was computed.  For each component of
$\phi$,
the power in all
the modes with $k\equiv |\vec k|$ in bins of width $0.057a^{-1}$
were averaged.
Figure~1a shows the time evolution
of one component of the pion field.
The curves plotted are (top to bottom in the figure)
the average power in the modes in the momentum bins centered at
$ka=$~0.20, 0.26, 0.31, 0.37, 0.48, 0.60, 0.71, 0.94, 1.16, 1.39, and 1.84.
The initial power spectrum is white and all the
curves start at $t=0$ at approximately $0.01$ in lattice units.
Hence, the longest wavelength pion modes are amplified by a factor
of order 1000 relative to the shortest wavelength modes which
are not amplified at all.

Figure~1b shows the time evolution of the sigma field.
Only four modes are plotted --- $ka=$~0.20, 0.48, 0.94, and 1.84.
The vertical scale is different than in Figure~1a --- the sigma
modes do not grow nearly as much as the pion modes.}

\FIG\zzb{Time evolution of the spatially averaged
$\langle \phi^2 \rangle$ for the same simulation whose results
are shown in Figure~1.  The horizontal line is at
$\langle \phi^2 \rangle = v^2$.}

\FIG\zzc{Same as for Figure~1a, except that in this figure
we have used $a^{-1}=260~{\rm MeV}$ and have chosen
initial conditions from the distribution \bm\ following
the crude recipe outlined in the text to model a quench
from $T_i=1.2~T_c$ to $T=0$.  Note that the vertical
scale is different than in Figure~1a.  All the curves
start at about $0.03$ at $t=0$.
With these more
realistic initial conditions, long wavelength modes of
the pion field grow less than in Figure~1a,
but are still significantly amplified relative to shorter
wavelength modes.}

\endpage
\refout
\endpage
\figout
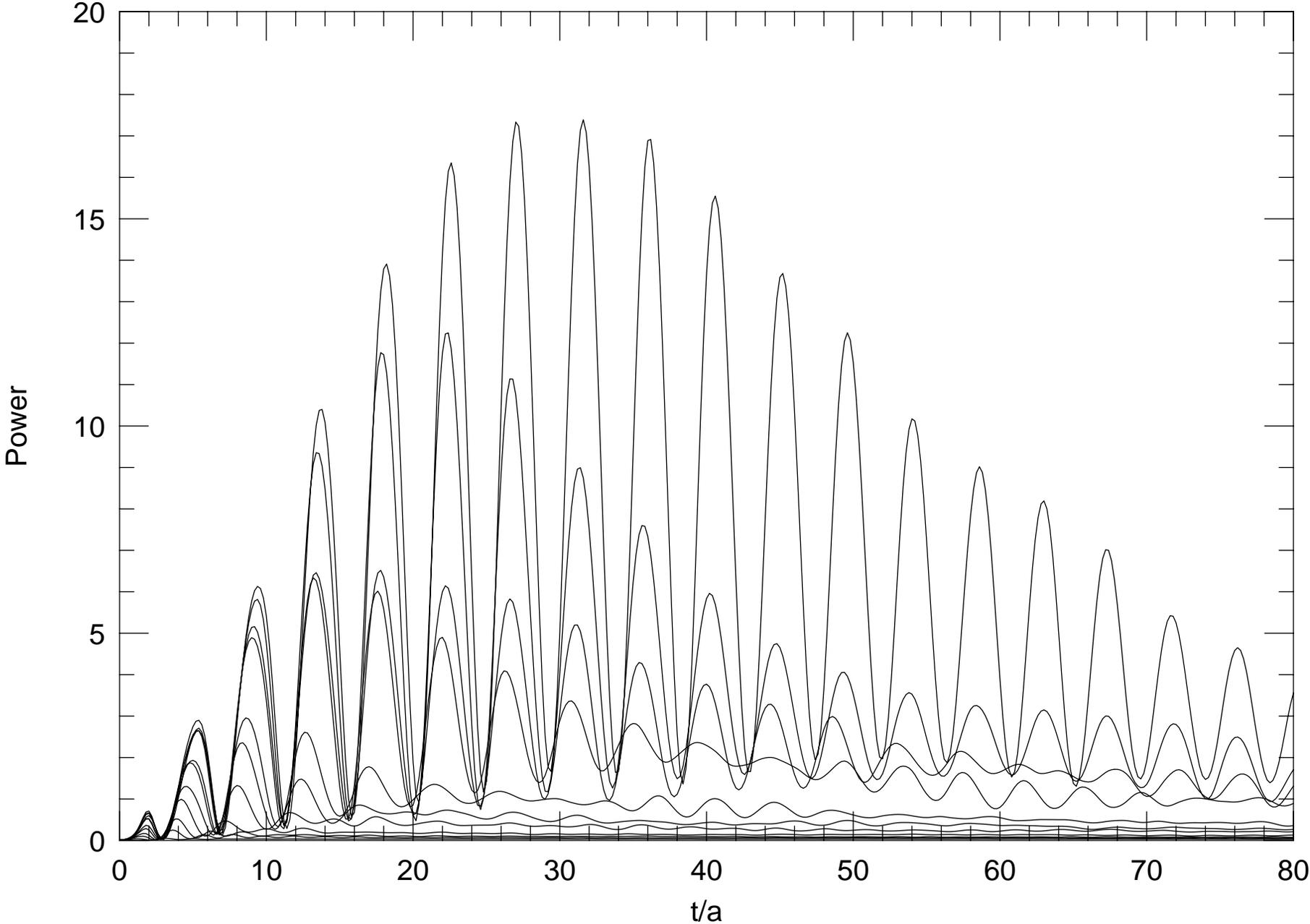
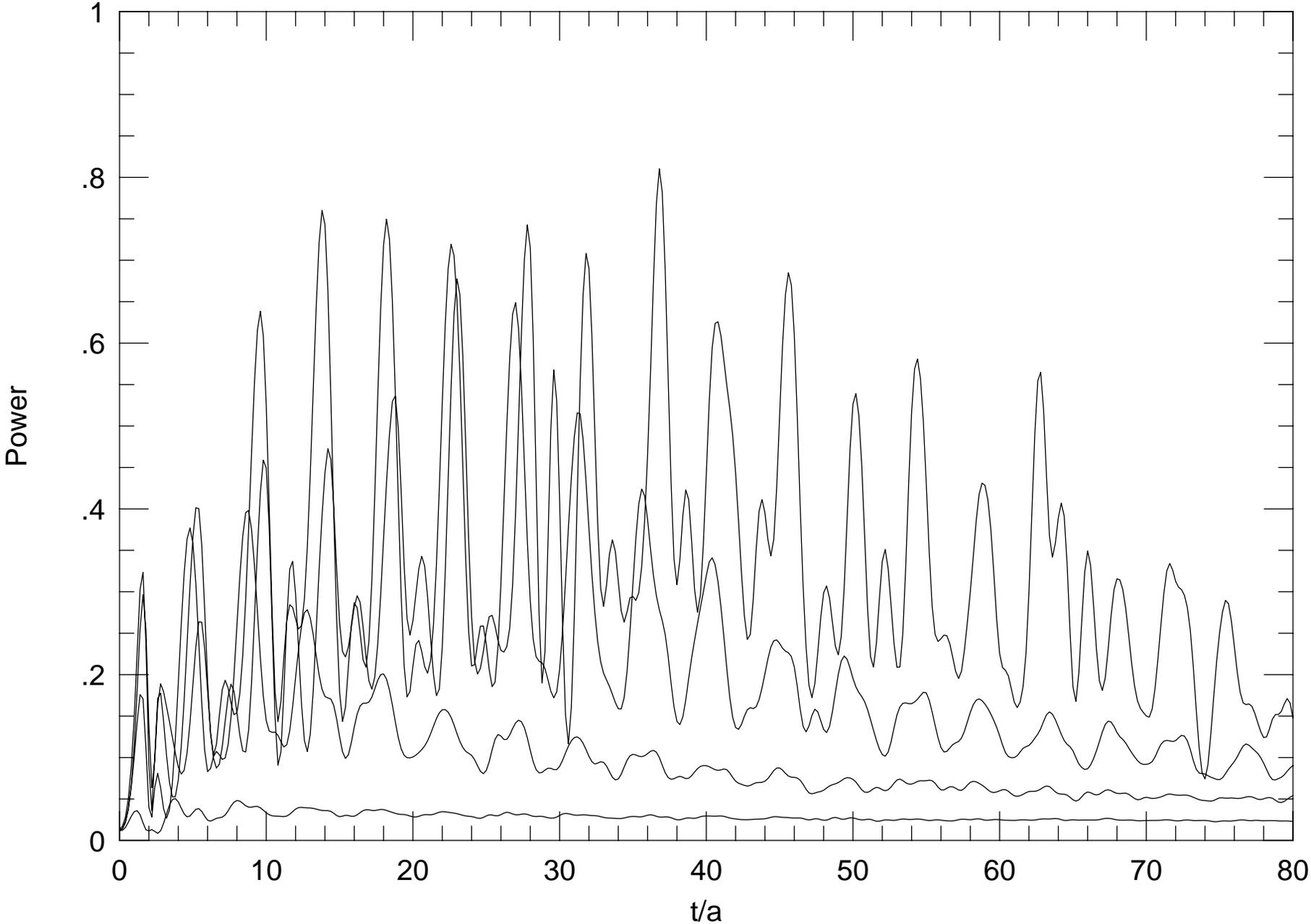
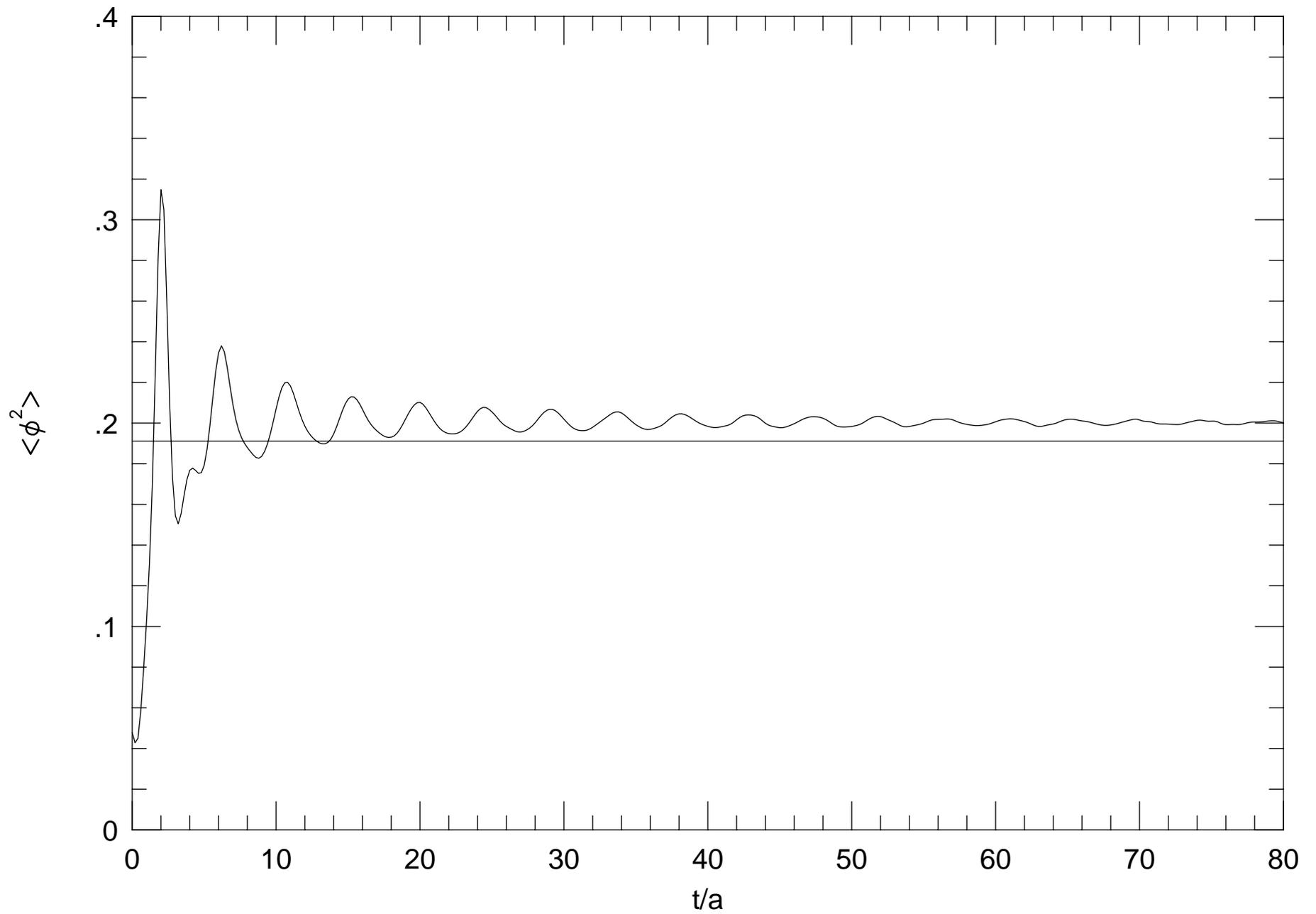
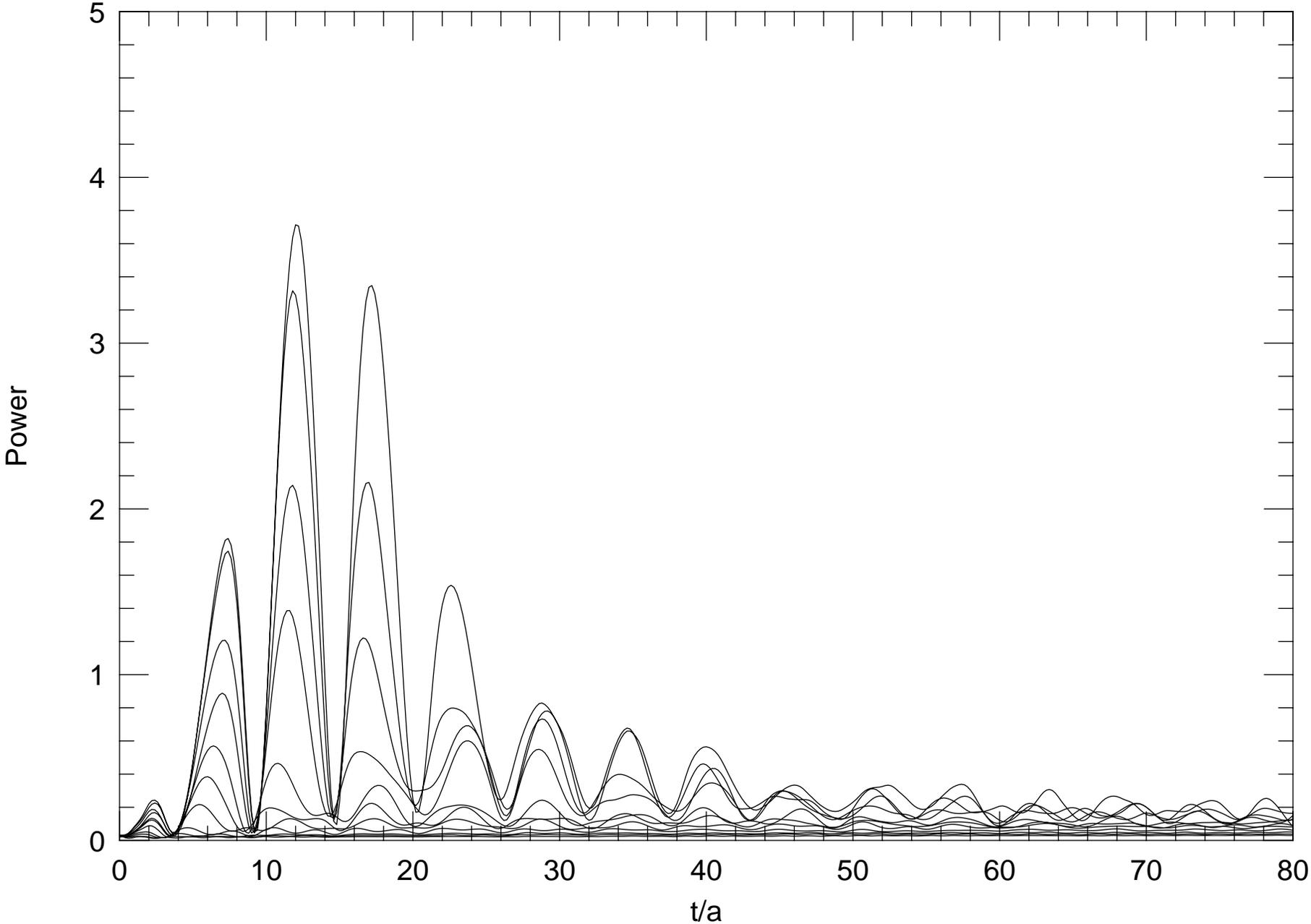
\end